\documentclass[a4paper,12pt]{article}
\usepackage[utf8]{inputenc}
\usepackage{cancel}
\usepackage{ulem}
\usepackage{amsfonts}
\usepackage{amssymb}
\usepackage{graphicx}
\usepackage{amsmath}
\usepackage{enumerate}
\usepackage{mathtools}
\usepackage{subfig}
\usepackage{color}
\usepackage{float}
\usepackage{here}
\usepackage{cite}
\usepackage{mathrsfs}
\usepackage{float,epsfig}
\usepackage{dcolumn}

\usepackage{graphicx}
\usepackage{bm}
\usepackage{amsmath,amssymb,amsthm}
\usepackage[colorlinks=true,linkcolor=blue,citecolor=red]{hyperref}
\makeatother

\begin{document}
\title{\normalsize
\phantom{fff}
\vspace{-3cm}
\begin{flushright}
FISPAC-TH/271/2020\\
UQBAR-TH/2020-091
\end{flushright}
\vspace{2cm}
{\bf \Large	Shadows   of  5D Black Holes from String Theory }}
\author{   \small A. Belhaj$^{1}$\footnote{belhajadil@fsr.ac.ma}, H. Belmahi$^{1}$,   M.  Benali$^{1}$,   W. El Hadri$^{1}$,  H. El Moumni$^{2}$\thanks{hasan.elmoumni@edu.uca.ma},
E. Torrente-Lujan$^{3}$\thanks{torrente@cern.ch}\footnote{ Authors in alphabetical order.}
	\hspace*{-8pt} \\
	{\small $^1$ D\'{e}partement de Physique, Equipe des Sciences de la mati\`ere et du rayonnement,
		ESMaR}\\ {\small   Facult\'e des Sciences, Universit\'e Mohammed V de Rabat,  Rabat, Morocco} \\
	{\small $^{2}$  EPTHE, Physics Department, Faculty of Science,  Ibn Zohr University, Agadir, Morocco}\\
 {\small $^{3}$ IFT, Dep. de F\'isica, Univ.  de Murcia,
Campus de Espinardo, E-30100 Murcia, Spain}\\
 }

 \maketitle

	\begin{abstract}
		{\noindent}
		
We  study   the   shadow  behaviors  of  five dimensional (5D) black  holes embedded in
type IIB superstring/supergravity inspired  spacetimes by  considering  solutions with and without rotations.
 Geometrical properties as  shapes and  sizes are  analyzed   in
terms of the D3-brane number and the rotation parameter. Concretely,  we   find that the shapes   are indeed significantly distorted by such physical parameters and 
the size of the shadows decreases with the brane or ``color''  number  and the  rotation. Then,  we investigate geometrical observables and  energy emission rate  aspects.

	\end{abstract}
\newpage

\tableofcontents

\section{Introduction}
Black holes 
have been extensively investigated in connections with many gravity theories.
Such investigations  have been boosted  by recent direct observations,
the   gravitational wave  events 
 and
 by the  first image provided by the Event Horizon telescope (EHT)  international collaboration \cite{1,2,3,4}.

Thermodynamical and optical aspects of numerous  black holes, in 4D and arbitrary dimensions, have been dealt with
using different approaches and methods including numerical ones \cite{5,6,7,8,9,10,11,110,12,13,15,150,151,152}.
A particular emphasis has been put on black hole objects living in higher dimensions involving  various  rotating parameters
required by space-time symmetries.  In such spaces, non-trivial extended  black objects can also  arise going
beyond black holes. The associated horizon  geometries have been obtained and  examined  with  using  alternative  ways.

Many  other  physical properties   have been shown in connections with non-trivial models of gravity in  extra dimensions.
 Specifically,  string theory motivated models have been used as playgrounds
  to study certain phase transitions of black holes in the presence of solitonic  D-brane objects.
  Precisely,  the Hawking-Page phase transition of black  holes in  the $AdS_{5}\times \mathbb{S}^{5}$ geometry have been treated  by considering the number of D3-branes as a thermodynamical variable \cite{16}.
  Such investigations  have  been  extended  to M-theory on $AdS_{p+2}\times \mathbb{S}^{11-(p+2)}$ geometries
  where $p=2$ and 5 \cite{17,18,19}.
  Motivations to consider such higher dimensional  frameworks  are  associated with  non-local objects  being useful to
   understand  the AdS/CFT conjecture. In  particular, 
    explicit  examples  of such a   correspondence are elaborated in many places including  \cite{ads1,ads2}.

More recently,  the shadows and other optical behaviors of the  black holes in higher dimensional space-times,  motivated by
supergravity models, have been  investigated. More precisely,  5D black holes involving  more than one rotating parameter
have been approached    using the shadow analysis. A special attention has been devoted to  Myers-Perry solutions\cite{20}.

The aim of this work is to  further advance in the study of properties as   the shadows of 5D black  holes embedded
in type IIB superstring geometries.  In particular, we consider  solutions without and with  rotations.
 Among other results, we  find that the shapes   are significantly distorted and the sizes are  decreasing with the
 increase of the  D3-brane number and the rotating parameter. The geometrical observables and the energetic  aspects are also  analyzed in some details.

This work is structured as follows. In section 2, we  give a fast  review on  5D black holes being embedded in  type IIB superstring theory. In section 3, we study the shadows of   non-rotating black holes in terms of the  D3-branes. In section 4, we  investigate    a  5D black hole with  a single rotating parameter. The geometrical observables and the  energetic emission aspects are analyzed in section 5. The last section is devoted to   conclusions  and open questions.

\section{5D rotating  type IIB stringy  black holes}

Let us consider 5D black  holes embedded in  a  type IIB superstring/supergravity geometry  with  D3-branes.
The physics of such solitonic objects  has been extensively studied  in connections  the 
AdS/CFT conjecture \cite{21,22,23}.
Properties like the thermodynamical aspects of 5D black holes in such backgrounds have been  dealt with by examining certain phase transitions including the Hawking-Page  one.
Here, we will be interested in  optical aspects  of such black holes in  the $AdS_{5}\times \mathbb{S}^{5}$ physical  space.  This
can be   considered  as the  near horizon geometry of   black D3-branes in ten dimensions.
Rotating  5D black hole solutions,   in general,  involve two rotation parameters  related to  the Casmir
operators of  the  $SO(5)$ Lie group.
 In this  way,  non-trivial optical properties depend  on such  parameters.
 Following to \cite{23,24}, the metric of a 5D   black hole  with two rotating parameters assumed to be 
  embedded in the $AdS_{5}\times \mathbb{S}^{5}$ geometry can be written, using
Boyer-Lindquist coordinates, as
\begin{eqnarray}
ds^2&=&-{\Delta_{r} \over{\rho}^2}
\left(dt-{a\sin^2{\theta}\over\Xi_a}d\phi-{b
\cos^2{\theta}
\over\Xi_b}d\psi\right)^2+{\rho^2\over\Delta_\theta}
d\theta^2+{\Delta_{\theta}\sin^2{\theta}\over\rho^2}\left(adt-{(r^2+a^2)
 \over {\Xi}_a} d\phi\right)^2\nonumber\\
&
&+{(1+r^2/\ell_{AdS}^2)\over r^2\rho^2}
\left(abdt-{b(r^2+a^2)
\sin^2\theta\over\Xi_a}d\phi-{a(r^2+b^2)
\cos^2\theta\over\Xi_b} d\psi\right)^2\nonumber\\
&
&+{\rho^2\over\Delta_{r}}dr^2+{\Delta_{\theta}\cos^2{\theta}
\over\rho^2}\left(bdt-{(r^2+b^2)\over\Xi_b}d\psi\right)^2 +\ell^{2}_{AdS} d\Omega_{5}^{2}, \label{A}
\label{a1000}
 \end{eqnarray}
where the involved terms are 
\begin{eqnarray}
\rho^{2}&=&r^2+a^2\cos^2\theta+b^2\sin^2\theta,\nonumber\\
 \quad \Xi_a&=&1-\frac{a^2}{\ell_{AdS}^2},
\quad \Xi_b=1-\frac{b^2}{\ell_{AdS}^2},\nonumber\\
\Delta_{r}&=&{1\over r^2}(r^2+a^2)(r^2+b^2)(1+\frac{r^2}{\ell_{AdS}^2})-m, \nonumber\\
\Delta_{\theta}&=&1-\frac{a^2}{\ell_{AdS}^2}\cos^2\theta-\frac{b^2}{\ell_{AdS}^2}\sin^2\theta.
\end{eqnarray}
The quantity $m$ is   a mass parameter and $a$ and $b$ are the two rotation parameters.  The two angular coordinates
$\phi$ and  $\psi$    are  constrained by  $0\leq\phi\leq2\pi$ and $0\leq\psi\leq\pi/2$.
$d\Omega_{5}^{2}$ is the metric of  the  five dimensional sphere $\mathbb{S}^{5}$.
From the  type IIB superstring/supergravity  point of view, the associated  ten dimensional factorized
  spacetime  can
be interpreted as the near horizon geometry of $N$ coincident configurations of the  D3-branes.
The AdS radius $\ell_{AdS}$ and the gravitational constant can be related to $N$  via the following relations
\begin{eqnarray}
\ell_{AdS}^{4}=\frac{\sqrt{2}N\ell_{p}^{4} }{\pi^{2}},\qquad
G_{5}=\frac{\ell_{p}^{8}}{\pi^{3}\ell_{AdS}^{5}}
= \frac{\ell_{p}^{3}}{2^{\frac{5}{8}}\sqrt{\pi}N^{\frac{5}{4}}}.
\label{ab}
\end{eqnarray}
It is   noted that the metric Eq.(\ref{A}) is singular when $g_{rr}=\Delta_r=0$ and $\rho^2=0$.
This  metric is invariant under the following symmetry
\begin{equation}
 a\leftrightarrow b, \quad   \theta\leftrightarrow\frac{\pi}{2}-\theta, \quad    \phi \leftrightarrow \psi.
\end{equation}
 For the sake of simplicity and to make contact with known results,  we will limit our analysis in this work to two cases:
   zero ($a=b=0$)  and one rotation parameter ($b=0$).
  A more  general   study will be the object of further investigations left  for future works.

\section{Shadows of  5D non-rotating black holes in presence of D3-branes}

Let us study the null geodesics (test ``photon'' orbits)   around such stringy type IIB  black holes.
We  consider first  a baseline case with no rotation. In this case   $a=b=0$, the   line element in Eq.(\ref{a1000}) reduces to
\begin{equation}
ds^{2}=-\Delta_rdt^{2}+\frac{1}{\Delta_r}dr^{2}+r^{2}\left( d\theta^{2}+\sin ^{2}\theta d\phi^{2}+\cos^{2}\theta d\psi^{2}\right)+\ell^{2}_{AdS} d\Omega_{5}^{2},
\label{1}
\end{equation}
where $\Delta_r$ is a  redefined radial function  given by  
\begin{equation}
\label{f}
\Delta_r=1-\frac{m}{r^2}+\frac{r^2}{\ell_{AdS}^{2}}.
\end{equation}
The  integration constant $m$  is related to the  black hole mass $M$  through the  relation
\begin{equation}
\label{m}
m=\frac{8}{3\pi} G_5 \times M,
\end{equation}
where  $G_5$ is  the  5D  gravitational constant. It is convenient  to    consider a   normalized  mass parameter $m'=\frac{3\pi}{4}m$. In this way, the above metric becomes
\begin{equation}
\Delta_r=1-\frac{2G_{5}M}{r^{2}}+\frac{r^{2}}{\ell^{2}_{AdS}}.
\label{2}
\end{equation}
In order to implement the D3-brane effect,   Eq.(\ref{2})  can be rewritten as (using Eq.(\ref{ab}))
\begin{equation}
\Delta_r=1-\frac{2^{\frac{3}{8}}M\ell^{3}_{p}}{N^{\frac{5}{4}}\sqrt{\pi}r^{2}}+\frac{\pi r^{2}}{2^{\frac{1}{4}}N^{\frac{1}{2}}\ell^{2}_{p}}.
\label{5}
\end{equation}
The ``photon'' orbits or null geodesics  of the  5D non-rotating  black hole geometry can be computed
 by  using   an  effective Hamilton-Jacobi approach
(see \cite{25,26}, for example, for more  detailed calculations).
 The corresponding Hamilton-Jacobi equation  for our case  is 
\begin{equation}
\frac{1}{2}g^{\mu\nu}\frac{dS}{dx^{\mu}}\frac{dS}{dx^{\nu}}+\frac{dS}{d\tau}=0,
\label{12}
\end{equation}
where $\tau$ is an  affine parameter, and where  $S$ is the Jacobian action being given by
\begin{equation}
S=-Et+L_{\phi}\phi+L_{\psi}\psi+S_{r}(r)+S_{\theta}(\theta).
\label{13}
\end{equation}
 $E$, $L_{\phi}$ and  $L_{\psi}$  are conserved quantities corresponding  to the energy and
the angular momentums of the ''photon'', respectively \cite{25}.
  It is noted that $S_r$ and $S_\theta$  are  functions of $r$ and  $\theta$  only, respectively.

\begin{figure}[t] 
		\begin{center}
		\centering
			\begin{tabbing}
			\centering
			\hspace{8.cm}\=\kill
			\includegraphics[scale=.5]{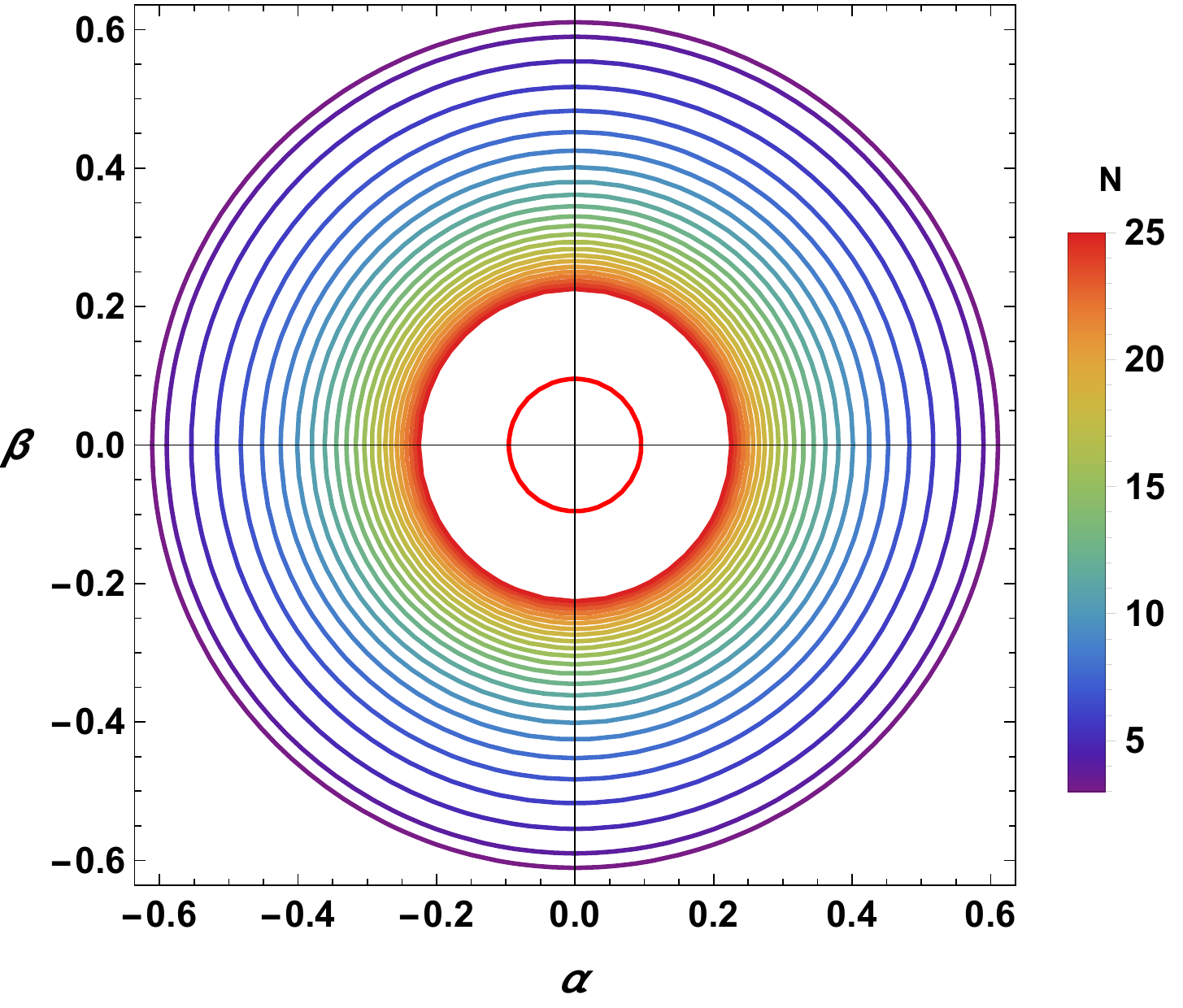} \>
			\includegraphics[width=8cm, height=6.3cm]{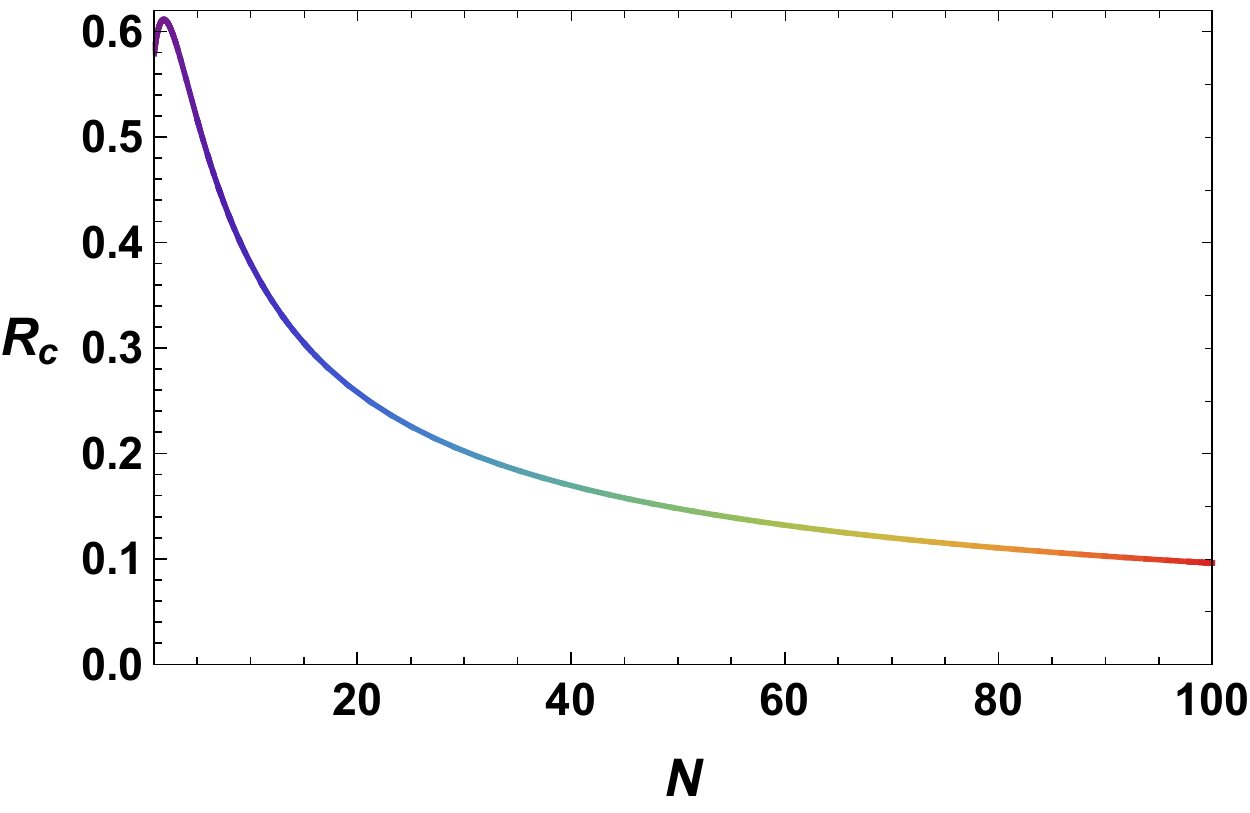} \\
		   \end{tabbing}
\caption{{
 {
  Left:} Black hole shadow in the celestial plane $(\alpha-\beta)$ for different values of  the brane number $N$  and
$\ell_{p}= 1,M=1$.
{
Right:} The variation of the shadow radius in terms of the branes number $N$.
In the left  panel, the red circle corresponds to  the brane number (N = 100). }}
\label{Smsh}
\end{center}
\end{figure}

Let us use a variable  separation method to solve the equation (\ref{12}).
Although the use of this method is not strictly necessary in the non-rotating case, it is convenient here to make contact with
the rotating case.
 Separating  functions of $r$ and $\theta$, from  Eq.(\ref{12}) and  Eq.(\ref{13}),   we get
\begin{eqnarray}
r^{2}\Delta_r\left( \frac{dS_{r}(r)}{dr}\right) ^{2}-\frac{r^{2}E^{2}}{\Delta_r}+L_{\psi}^{2}+L_{\phi}^{2}&=&-\mathcal{K}\\
\left( \frac{dS_{\theta}(\theta)}{d\theta}\right) ^{2}+L_{\phi}^{2}\cot ^{2}\theta+L_{\psi}^{2}\tan ^{2}\theta&=&\mathcal{K},
\end{eqnarray}
where $\mathcal{K}$ is  a  (real) constant  being  interpreted  as   the Carter constant and 
 considered as  an additional constant of motion \cite{26}. According to the Hamilton-Jacobi method, we find   the following set of equations
\begin{eqnarray}
\frac{dt}{d\tau}&=&\frac{E}{\Delta_r},\\
r^{2}\frac{dr}{d\tau}&=& \pm \sqrt{\mathcal{R}} ,\\
r^{2}\frac{d\theta}{d\tau}&=& \pm \sqrt{\Theta},\\
\frac{d\phi}{d\tau}&=&\frac{L_{\phi}}{r^{2}\sin ^{2}\theta},\\
\frac{d\psi}{d\tau}&=&\frac{L_{\psi}}{r^{2}\cos ^{2}\theta},
\end{eqnarray}
where  the  terms $\mathcal{R}$ and $\Theta$ are given, respectively,  by
\begin{eqnarray}
\mathcal{R}&=& -\Delta_rr^{2}\left(\mathcal{K}+L_{\phi}^{2}+L_{\psi}^{2}\right) +E^{2}r^{4} \\
\Theta &=& \mathcal{K}- L_{\phi}^{2}\cot^{2}\theta-L_{\psi}^{2}\tan^{2}\theta.
\end{eqnarray}
To  get  the shadow boundaries, the expression of the effective potential $V_{eff}$ for a radial motion will be  needed.  Using the relation
\begin{equation}\left( \frac{dr}{d\tau}\right) ^{2}+V_{eff}=0,\end{equation}
we obtain  the effective potential in such a  type IIB stringy  solution
\begin{equation}
V_{eff}=\frac{\Delta_r}{r^{2}}\left(\mathcal{K}+L_{\phi}^{2}+L_{\psi}^{2}\right)-E^2.
\end{equation}
The unstable circular orbits  relying on  the maximum of the effective potential are  determined by imposing the following constraints
\begin{equation}
V_{eff}=\frac{dV_{eff}}{dr}\Bigm |_{r=r_{0}}=0, \qquad \mathcal{R}=\frac{d\mathcal{R}}{dr}\Bigm |_{r=r_{0}}=0.
\label{26}
\end{equation}
 Eq.(\ref{5}) can be developed  to provide
\begin{equation}
\frac{dV_{eff}}{dr}\Bigm|_{r=r_{0}}=-2 \frac{\left(\mathcal{K}+L_{\phi}^{2}+L_{\psi}^{2}\right)\left(N^{\frac{5}{4}}\sqrt{\pi}r^{2}-2^{\frac{11}{8}}M\ell_{p}^{3}\right)}{N^{\frac{5}{4}}\sqrt{\pi}r^{5}}=0,
\end{equation}
where  $r_{0}$ is the unstable photon orbit radius. The calculation gives
\begin{equation}
r_{0}=\sqrt{\frac{2^{\frac{11}{8}}M\ell^{3}_{p}}{N^{\frac{5}{4}}\pi^{\frac{1}{2}}}}.
\label{f}
\end{equation}
Introducing  the  impact parameters
\begin{equation}
\xi_{1}=\frac{L_{\phi}}{E}\qquad \xi_{2}=\frac{L_{\psi}}{E} \qquad \eta =\frac{\mathcal{K}}{E^{2}},
\end{equation}
we find the following  algebraic geometrical relation
\begin{equation}
E^{2} {r_0^{2}- E^{2} \left( \eta+\xi^{2}_{1}+\xi_{2}^{2}\right) \left(1-\frac{2^{\frac{3}{8}}M\ell^{3}_{p}}{N^{\frac{5}{4}}\sqrt{\pi}r_0^{2}}+\frac{\pi r_0^{2}}{2^{\frac{1}{4}}N^{\frac{1}{2}}\ell^{2}_{p}}\right)}=0.
\end{equation}
Using Eq.(\ref{f}), this   relation  can be reformulated as follows
\begin{equation}
\eta+\xi_{1}^{2}+\xi_{2}^{2}=\frac{2^{\frac{19}{8}}M\ell^{3}_{p}\sqrt{N}}{N^{\frac{7}{4}}\sqrt{\pi}+2^{\frac{17}{8}}\pi M \ell_{p}},
\end{equation}
being a two-dimensional   shadow geometry.  However, to visualise  the corresponding   black hole  shadows, we introduce  the  celestial coordinates \cite{Amir:2017slq}. Taking
the limit of a far away observer, we use  such 
coordinates, considered  as a function of the constant of motion
\begin{eqnarray}
\alpha &=&\lim\limits_{{r_{O}}\rightarrow +\infty} -r_{O}^{2}\left(\sin\theta\frac{d\phi}{dr}+\cos\theta\frac{d\psi}{dr}\right),\\
\beta&=&\lim\limits_{{r_{O}}\rightarrow +\infty}r_{O}^{2}\frac{d\theta}{dr},
\end{eqnarray}
where $r_{O}$ denotes the distance between   the black hole and the observer.  Using the expressions
\begin{eqnarray}
\frac{dr}{d\tau}=E\sqrt{1-\frac{\Delta_r\left( \eta+\xi_{1}^{2}+\xi_{2}^{2}\right) }{r^{2}}}\label{j},\qquad
\frac{d\theta}{d\tau}=\frac{\sqrt{\mathcal{K}-L_{\phi}^{2}\cot^{2}\theta-L_{\psi}^{2}\tan^{2}\theta}}{r^{2}}\label{k},
\end{eqnarray}
 we get, for these coordinates,  the relations
\begin{eqnarray}
\alpha &=&- \left( \frac{\xi_{1}}{\sin \theta}+\frac{\xi_{2}}{\cos \theta}\right) , \\
\beta &=&  \pm \sqrt{\eta-\xi_{1}^{2}\cot ^{2}\theta-\xi_{2}^{2}\tan ^{2}\theta}.
\end{eqnarray}
Let us study some   particular solutions for the sake of illustration.
For instance,   we can consider the case when an observer is situated in the equatorial hyperplane
 in 5D, where  the inclination angle is $\theta_{0}=\frac{\pi}{2}$ and $\xi_{2}=0$.
 Concretely, the   shadow geometry  for a 5D non-rotating  black hole,  in the presence of $N$ D3-branes,  is
 characterized by the relation
\begin{equation}
\alpha^{2}+\beta^{2}=\eta+\xi_{1}^{2}=\frac{2^{\frac{19}{8}}M\ell^{3}_{p}\sqrt{N}}{N^{\frac{7}{4}}\sqrt{\pi}+2^{\frac{17}{8}}\pi M \ell_{p}}.
\end{equation}
To  inspect   the D3-brane  effect,
 we  discuss   the  shadow  behaviors  and their   radius  $R_c$   as a function of the  D3-brane number $N$.
 An examination shows that the D3-brane number can be considered a real parameter controlling
the involved size.
It can be seen clearly that the  shadow shape is circular, while the
size decreases by increasing the  brane number  from a maximum value
\begin{equation}
R^{max}_{c}=\frac{2 \sqrt[7]{2}\ 5^{5/7} M^{2/7} \ell_p ^{16/7}}{7 \pi ^{6/7}},\quad
\end{equation}
at
\begin{equation}
\quad N^{max}= \frac{2\ 2^{11/14} \pi ^{2/7} M^{4/7} \ell_p ^{4/7}}{5^{4/7}}.
\end{equation}
  The radius decreases with the   D3-brane number and it goes to zero  in large $N$ limit. In particular,  this could also
  be  seen directly from the metric function $\Delta_r$.  This dependence on the brane number is similar to the one  of  black hole properties obtained recently  in M-theory scenarios \cite{27}.
We present a illustration of the results in this non-rotating case    in Fig.\ref{Smsh}.

Having examined the non-rotating black hole shadows with parallel  D3-brane configurations,
we will  focus in the next section on a  more general case by taking into account the rotation effect.

\section{{Shadows of 5D rotating  black holes}}

We go now beyond the previous results by  implementing  a single rotating  parameter.
Taking $b=0$ for simplicity, the line element  of  a  5D black  solution  reduces to
\begin{eqnarray}
ds^2&=&-\frac{\Delta_{r}}{\rho^2}\left( dt-\frac{a}{ \Xi_a}\sin^{2}\theta d\phi\right) ^{2}+\rho^2\left( \frac{dr^2}{\Delta_{r}}+\frac{d\theta^{2}}{\Delta_{\theta}}\right) +r^{2}\cos^{2}\theta d\psi^{2}\\
&+&\frac{\Delta_{\theta}\sin^{2}\theta}{\rho^2}\left( adt - \frac{r^{2}+a^{2}}{ \Xi_a}d\phi\right) ^{2} +\ell^{2}_{AdS} d\Omega_{5}^{2},\notag
\end{eqnarray}
where one has now  the  following reduced terms 
\begin{eqnarray}
\Delta_{r}&=&( r^{2}+a^{2})( 1+\frac{r^{2}}{\ell^{2}_{Ads}}) -m, \qquad
\Delta_{\theta}=1-\frac{a^{2}}{\ell_{AdS}^{2}}\cos^{2}\theta,\\
\rho^2&=& r^{2}+a^{2}\cos^{2}\theta, \qquad \qquad \Xi_a =1-\frac{a^2}{\ell^2_{AdS}}.
\end{eqnarray}
Using  the same  Hamilton-Jacobi formalism as in the previous section, we get the following  variable separated
 relations
\begin{eqnarray}
\Delta_{r}\left( \frac{dS_{r}}{dr}\right) ^{2}-\frac{\left[ E(a^{2}+r^{2})-aL_{\phi}\Xi_a\right] ^{2}}{\Delta_{r}}+\frac{\left( L_{\phi} \Xi_a-aE\right) ^{2}}{\Delta_{\theta}}+L_{\psi}^{
2}\left( 1+\frac{a^{2}}{r^{2}}\right)& =&-\mathcal{K} \\
\Delta_{\theta}\left( \frac{dS_{\theta}}{d\theta}\right) ^{2}+\cos^{2}\theta\left( \frac{L_{\phi}^{2}\Xi_a^{2}}{\sin^{2}\theta}-a^{2}E^{2}\right) +L_{\psi}^{2}\tan^{2}\theta&=&\mathcal{K},
\end{eqnarray}
where the notation is the same as in the previous section.
Then, we obtain  the geodesic equations
\begin{eqnarray}
\rho^2\frac{dt}{d\tau}&=&\frac{(r^{2}+a^{2})\left[ E(r^{2}+a^{2})-aL_{\phi}\Xi_a\right]}{ \Delta_{r}}+\frac{a(L_{\phi}\Xi_a -aE \sin^{2}\theta)}{\Delta_{\theta}},\\
\rho^2\frac{dr}{d\tau}&=&\pm\sqrt{\mathcal{R}},\\
\rho^2\frac{d\theta}{d\tau}&=& \pm \sqrt{\Theta},\\
\rho^2\frac{d\phi}{d\tau}&=&\Xi_a\left[ \frac{Ea(r^{2}+a^{2})-a^{2}L_{\phi}\Xi_a}{\Delta_{r}}+\frac{Ea\sin^{2}\theta-L_{\phi}\Xi_a}{\sin^{2}\Delta_{\theta}}\right],\\
\frac{d\psi}{d\tau}&=&\frac{L_{\psi}}{r^{2}\cos^{2}\theta},
\end{eqnarray}
where  $\mathcal{R}$ and $\Theta $ are  now  given by
\begin{eqnarray}
\Theta &=& E^{2}\left[ \Delta_{\theta}\left( \eta-\xi_{2}^{2}\tan^{2}\theta\right) -\cos^{2}\theta\left( \frac{\xi_{1}^{2}\Xi_a^{2}}{\sin^{2}\theta} -a^{2}\right) \right] ,\\
\mathcal{R}&=& E^{2}\left[ \left( a^{2}+r^{2}-a\xi_{1}\Xi_a\right)^{2}-\Delta_{r}\left( \frac{\left( a-\xi_{1}\Xi_a\right) ^{2}}{\Delta_{\theta}}+\eta+\frac{(r^{2}+a^{2})}{r^{2}}\xi_{2}^{2}\right) \right].
\end{eqnarray}
For $\theta=\frac{\pi}{2}$ and $\psi=\frac{\pi}{2}$ requiring  $\xi_{2}=0$, we  find  immediately  the   impact parameters
\begin{eqnarray}
\eta &=& \frac{r^{2} \left[ 16 a^{2}\Delta_{r}-\left( r\Delta'_{r}-4\Delta_{r}\right) ^{2}\right] }{a^2\Delta_{r}'^2}\vert_{r=r_{0}},\\
\xi_{1}&=& \frac{(a^{2}+r^{2})\Delta_{r}'-4r\Delta_{r}}{a\Delta_{r}'\Xi_a}\vert_{r=r_{0}},
\end{eqnarray}
where we have used  $\Delta'_{r}=\frac{\partial \Delta_{r}}{\partial r}$. Exploiting  the previous expressions,  these  impact parameters take the following form
\begin{eqnarray}
\eta&=&\frac{4 \mu  \left(a^2-\mu \right)-r^4\left(a^2 \nu -1\right)^2+4 r^2 \left(a^2 \mu  \nu +\mu \right)}{\left(a^3 \nu +2 a \nu  r^2+a\right)^2},\label{e2001}\\
\xi_{1}&=&\frac{a^4 \nu +a^2 \nu  r^2-a^2+2 \mu -r^2}{a^3 \lambda  \nu +a \lambda +2 a \lambda  \nu  r^2},\label{e2002}
\end{eqnarray}
where $\mu$, $\nu$, and $\lambda$ are given by
\begin{equation}
\mu =\frac{2^{3/8} \,\ell_p^3 \, M}{\sqrt{\pi }  \,n^{5/4}}, \hspace{1cm} \nu =\frac{\pi }{{2}^{1/4}  \, \ell_p^2  \,{n}^{1/2}},  \hspace{1cm}  \lambda =1-\frac{\pi  a^2}{{2}^{1/4} \, \ell_p^2 \, {n}^{1/2}}.
\end{equation}
To visualise the associated shadows, we need to introduce  the  celestial coordinates.
For the present model, they  are modified   as follows
\begin{eqnarray}
\alpha &=&-\left( \frac{\xi_{1}}{\sin \theta}+\frac{\xi_{2}}{\cos \theta}\right) ,\\
\beta &=& \pm \sqrt{\eta-\xi_{1}^{2}\cot^{2}\theta-\xi_{2}^{2}\tan^{2}\theta+ a^{2}\cos^{2}\theta}.
\end{eqnarray}
In the equatorial plane $\theta = \frac{\pi}{2}$, these  celestial coordinates  reduce to
\begin{eqnarray}
\alpha &=&-\xi_{1},\\
\beta &=&\pm\sqrt{\eta}.
\end{eqnarray}

\begin{figure}[!htb]
		\begin{center}
		\centering
			\begin{tabbing}
			\centering
			\hspace{8.6cm}\=\kill
			\includegraphics[scale=.5]{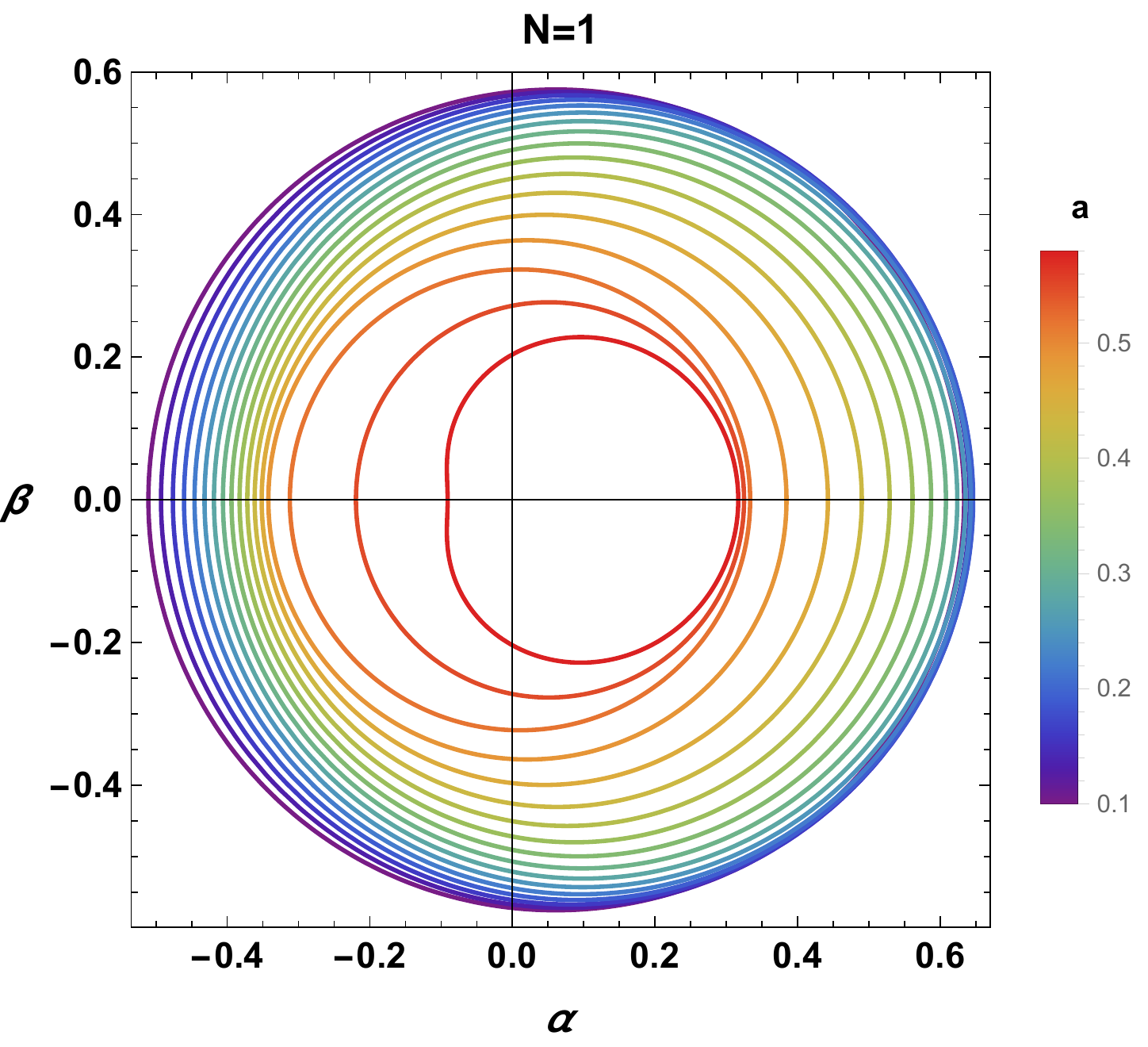} \>
			\includegraphics[scale=.5]{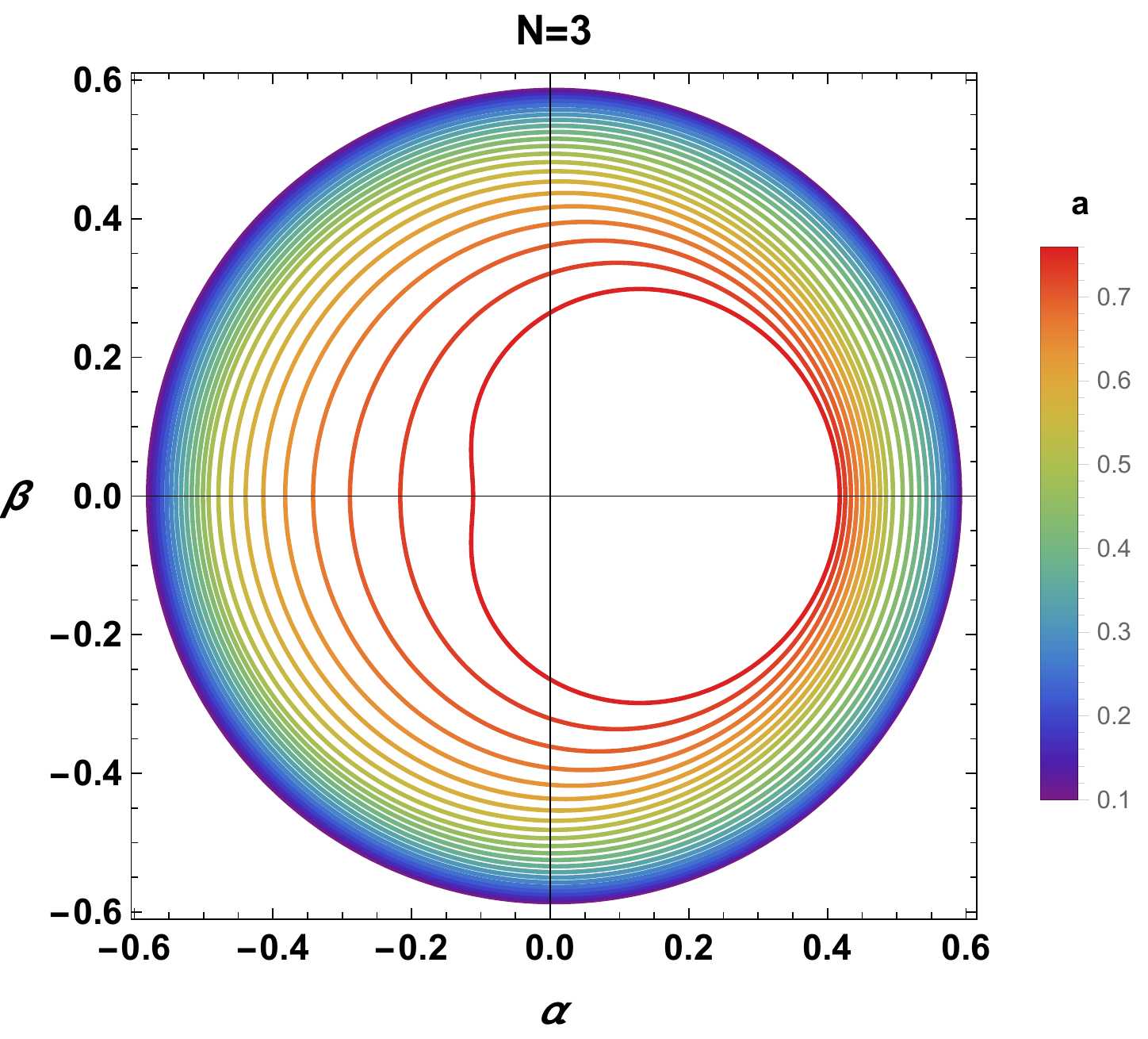} \\
			\includegraphics[scale=.5]{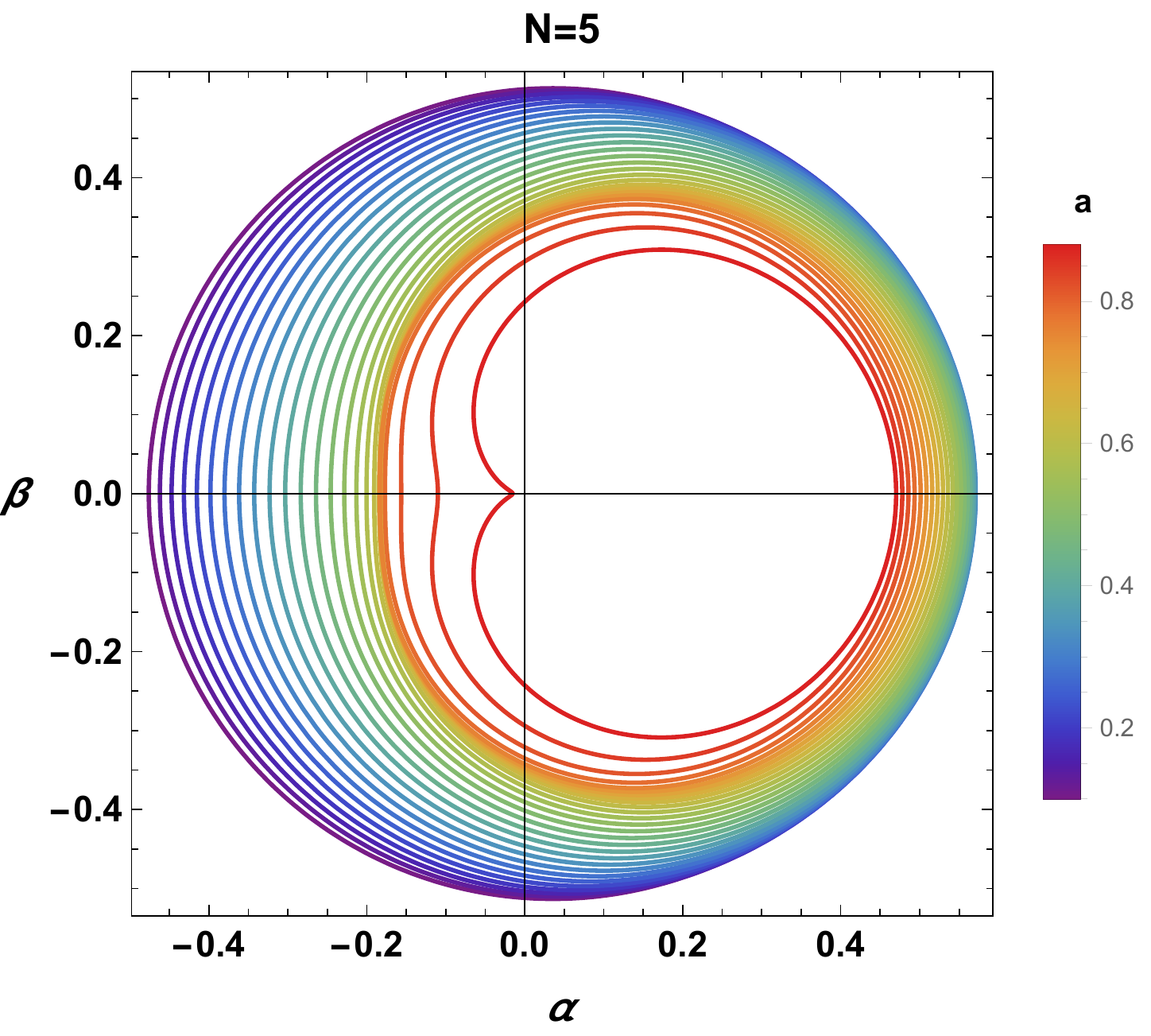} \>
			\includegraphics[scale=.52]{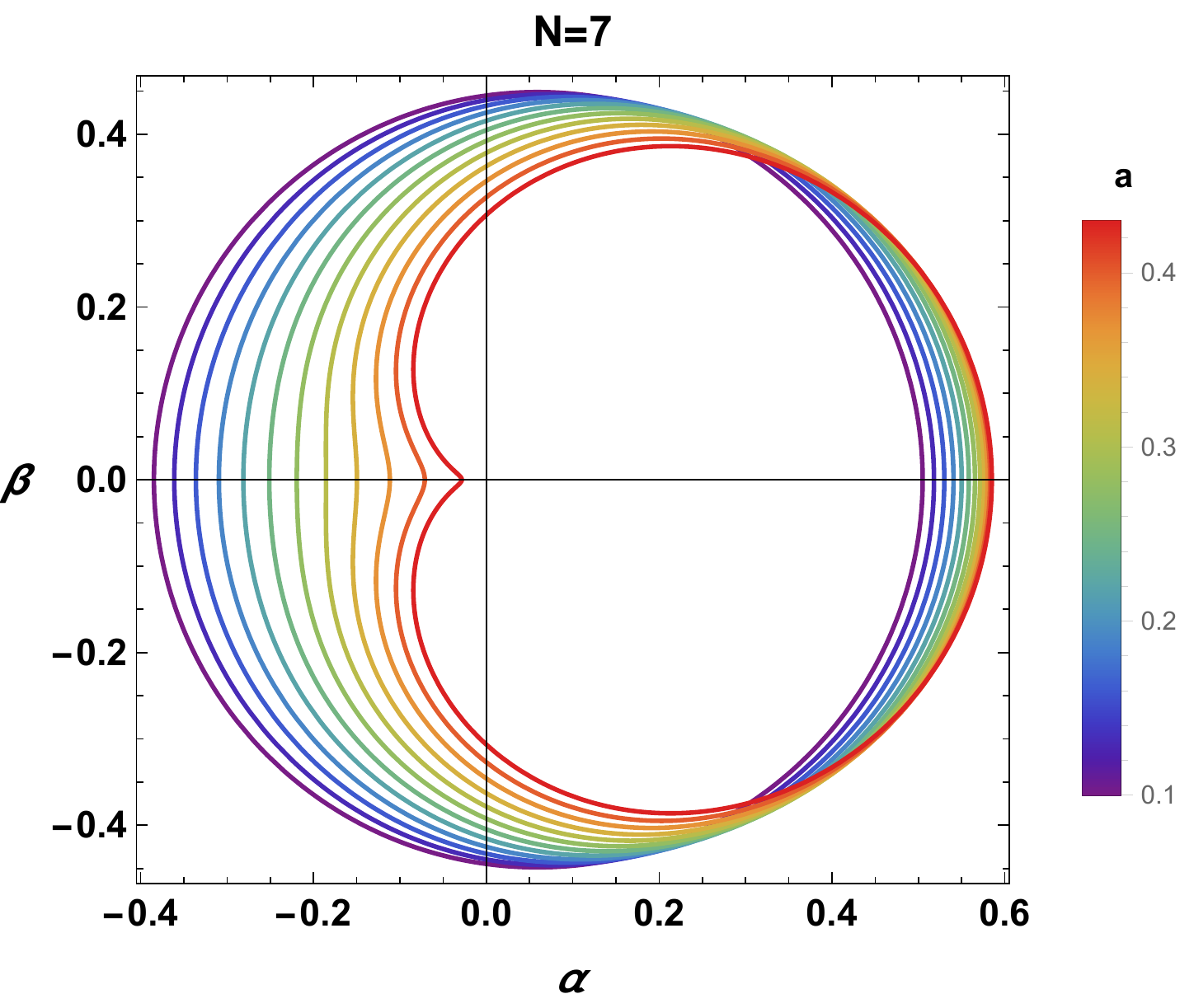} \\
		   \end{tabbing}
\caption{{
Shadow geometry of a rotating AdS black hole for different values of the  brane number $N$ and the  rotating rate $a$
(for $\ell_p=1$ and $M=1$). }}
\label{shfa}
\end{center}
\end{figure}

A close inspection of Eqs.(\ref{e2001},\ref{e2002}) reveals  that the D3-brane number impose
strong constraints on  the rotating parameter. The possible range of the  values of it  depends on such a number $N$.
 This behavior  is illustrated in Fig.\ref{shfa} where the parametric curves $(\alpha(r),\beta(r))$  are shown for a set of
fixed parameters $a$ and $N$. To get significant  shadows as  functions of  the rotation parameter $a$,  we should consider  $N$ lower  than a certain number (less than $\approx 8$ in the case of the illustrating  figure).

Fixing the D3-brane number, one observes that  the shadow size decreases by increasing the rotating parameter $a$.
For  $N=1$, the shadows keep  the  circular  behaviors for small values of $a$.  For $N>1$, however,  the shadow starts developing a D-form for the high values of the rotation parameter near $a\simeq 0.7$. As the number of the D3-branes increases the D-shape becomes of ``cardioid'' type.
Similar effects were   firstly  uncovered in \cite{150}.
This  complicated geometry has been shown to appear either  in M-theory scenarios   in the presence of the M2-branes \cite{27}
or in  charged rotating  black holes with a  cosmological constant \cite{150}.
Let us finally note that  the black hole shadow is controlled   basically by three parameters  $(M,N,a)$
namely  a moduli space. Not all regions in such a  space are acceptable.

 \section{Geometrical observable and energy emission rate aspects}

In  this section, we discuss further geometrical observables and  the  behavior of the black hole energy emission rate as a function of the  brane number and  the rotation.
The  observation of the shadow of the supermassive black hole  $M87^\star$ has opened a promising
  to probe gravity theories beyond GR, in particular candidates to quantum gravity as M-theory \cite{27} or string theory (as in the present work).
  We show next that the present  or near-future observational
data is indeed sensitive to stringy brane effects, i.e. through $N$, the number of D3-branes, and, it potentially could
 be used to put constraints on the  black hole parameters and then on the string theory itself.

\subsubsection*{Geometrical  physical observables}

\begin{figure}[t]   
		\begin{center}
		\centering
			\begin{tabbing}
			\centering
			\hspace{8.4cm}\=\kill
			\includegraphics[scale=.5]{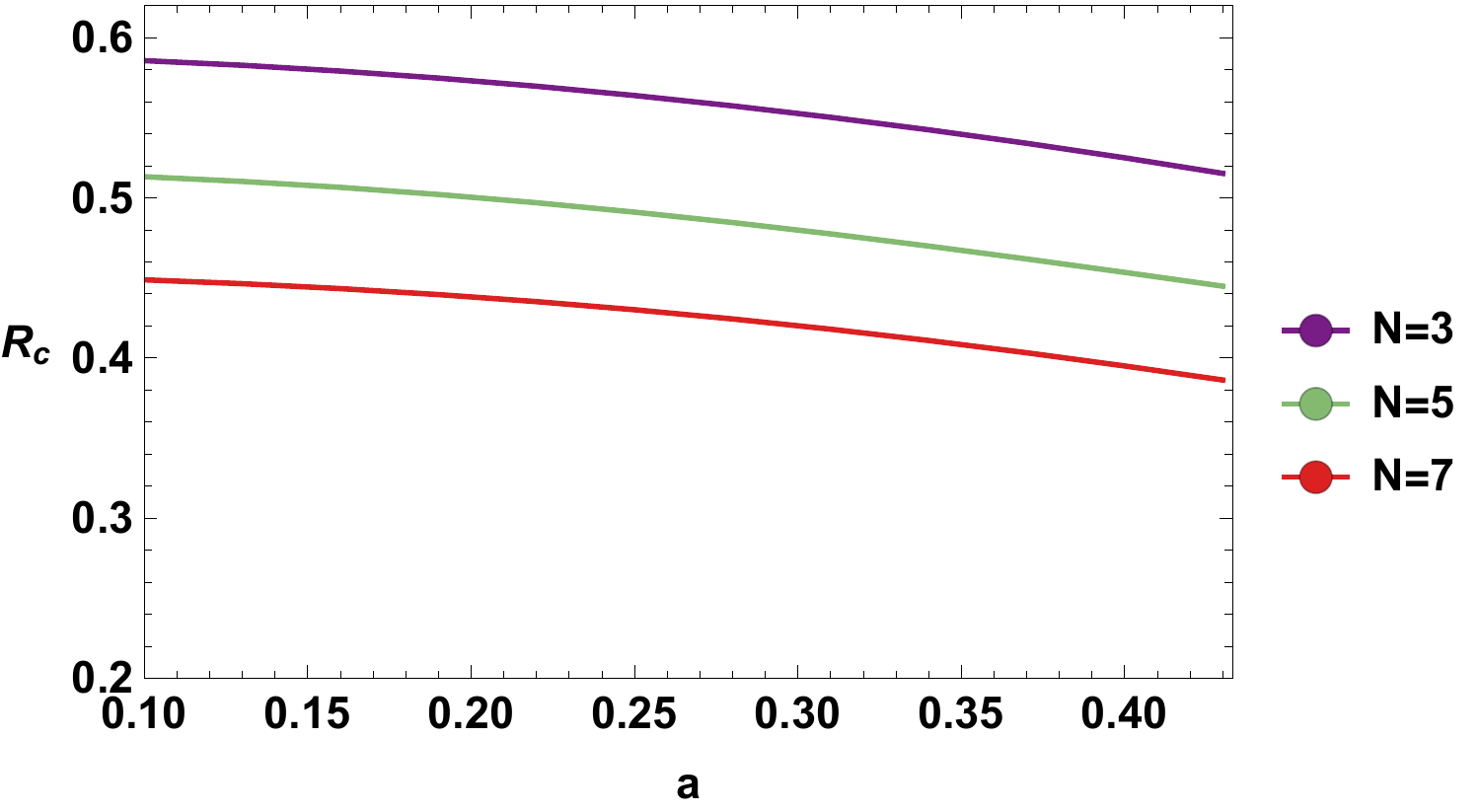} \>
			\includegraphics[scale=0.5]{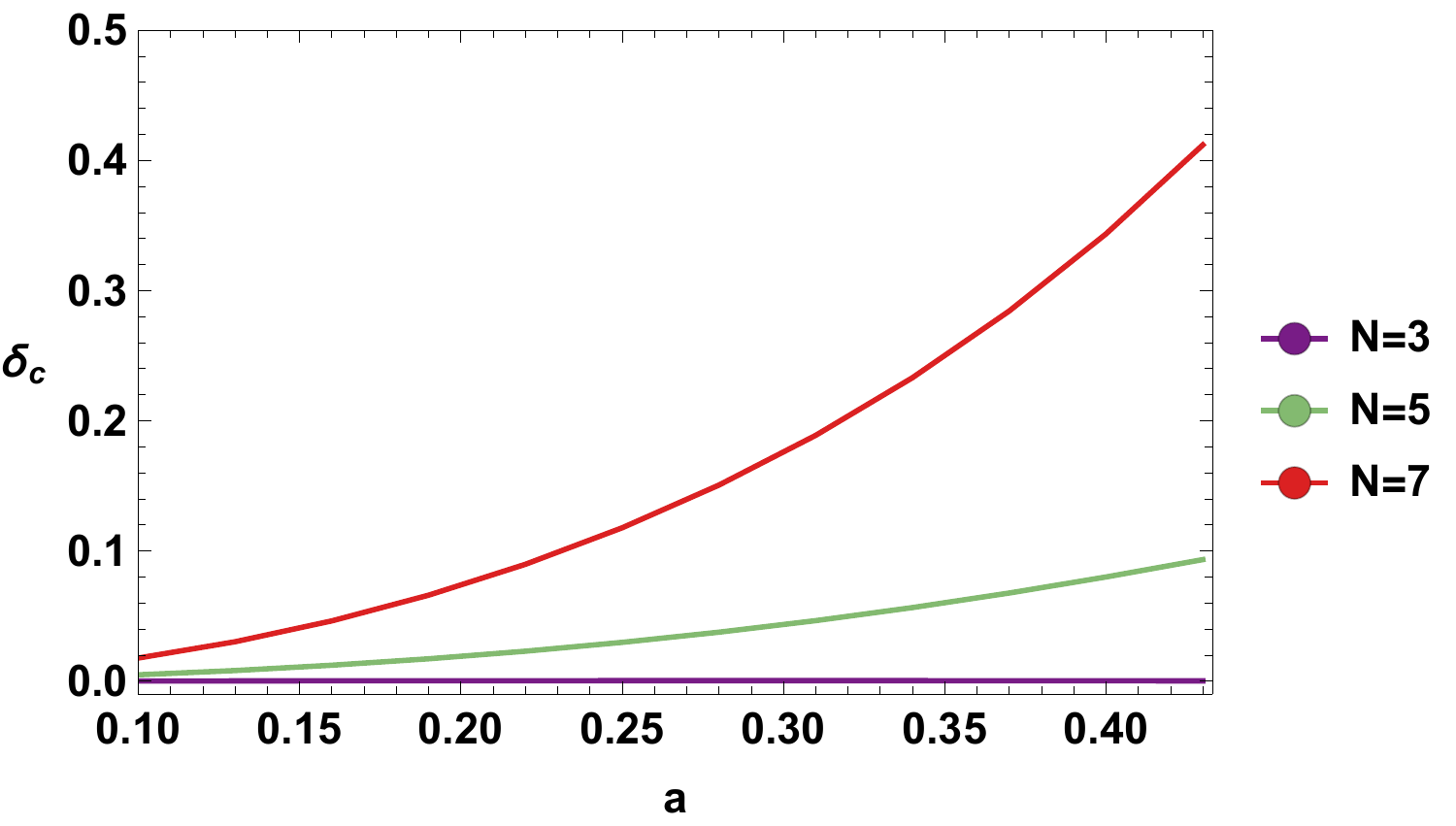} \\
			\includegraphics[scale=.5]{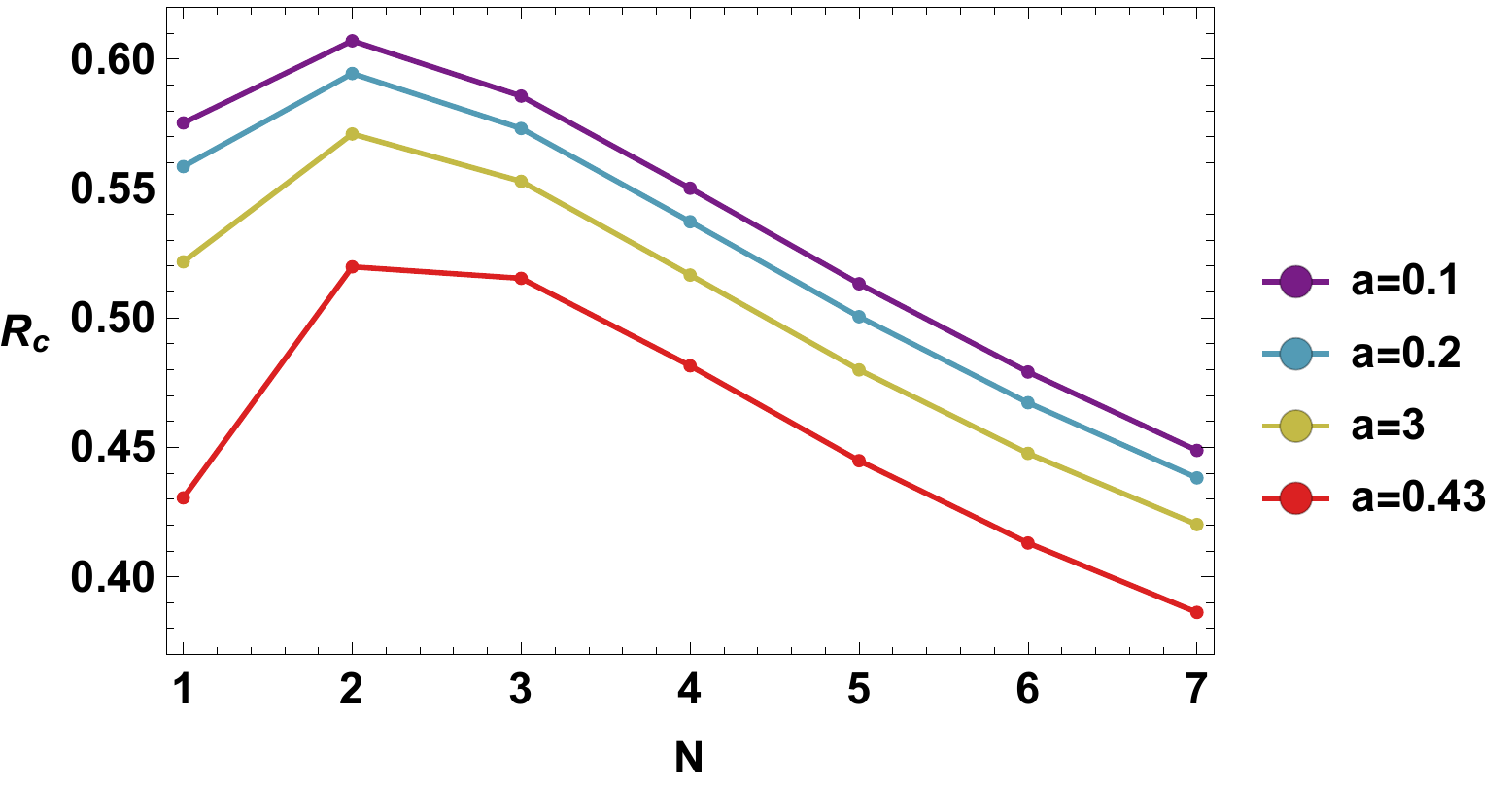} \>
			\includegraphics[scale=.52]{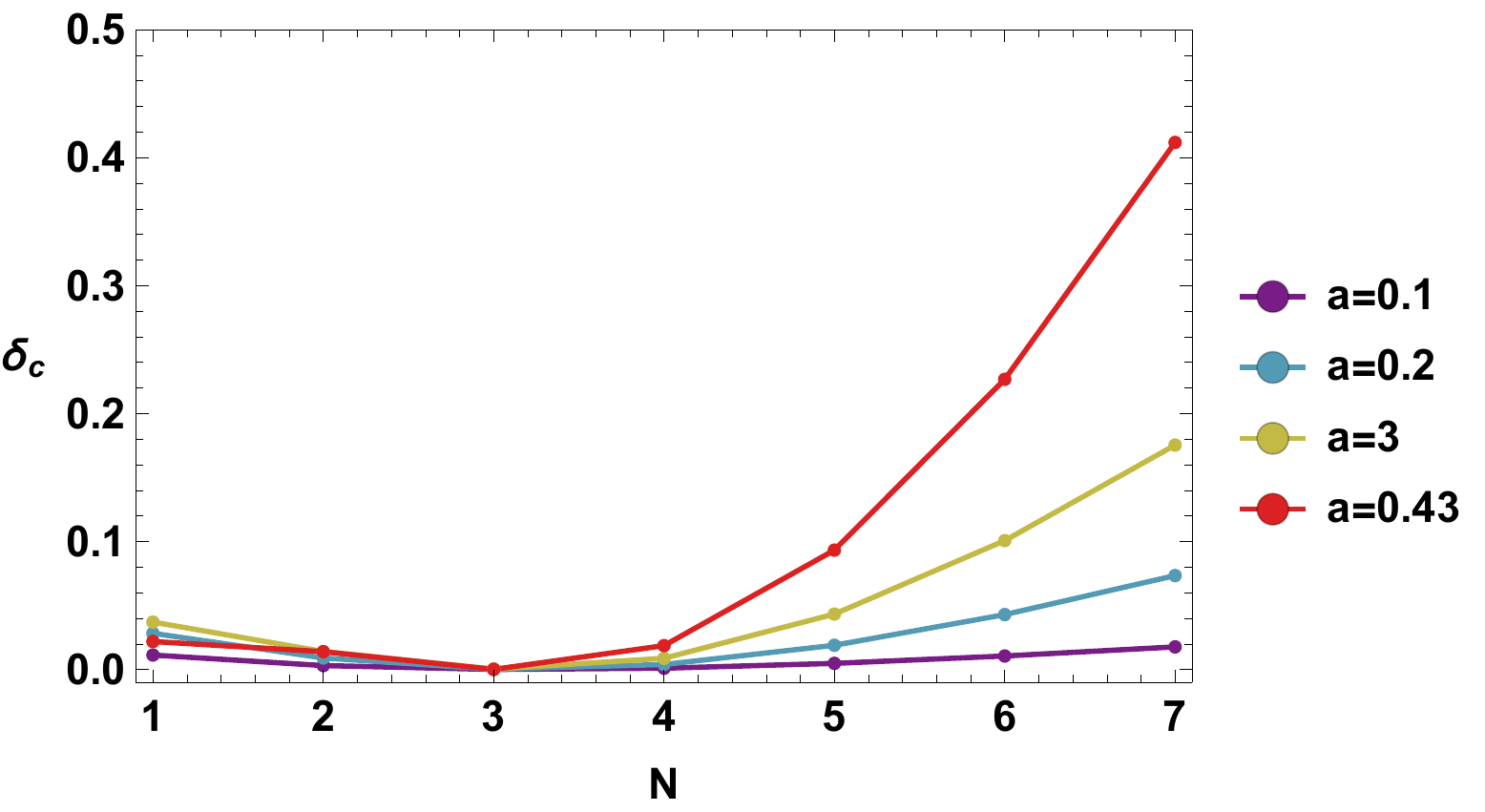}
		   \end{tabbing}
\caption{{
Astronomical observation variables
 for different values of the brane numbers $N$, the  rotating rate $a$,  for $\ell_p=1$ and $M=1$.}}
\label{diso}
\end{center}
\end{figure}

We approach now the study of observables
which can be useful for  making contact with present or future black hole observations.
To get  geometrical    physical data,   one needs  two
 observables.  They are  for example  the radius $R_c$ of
the shadow and the distortion $\delta$ controlling the size and shape respectively.
 Before going ahead, let us briefly  recall  the definition of them.
 To   facilitate    the discussion, we consider shadow configurations   with  $\xi_2=0$ as in the previous section.
   Following \cite{28},  the shadow of the black hole is characterized by three specific points.
   They are  the  top and the  bottom positions  ($\alpha_t,\beta_t$), ($\alpha_b,\beta_b$)  respectively, and  the point of a
   standard  circle ($\alpha_p,0$).
   The point of  the distorted shadow circle ($\alpha_p,0$) intersects the horizontal axis  at $\alpha$.
   The distance between the two  points   is   controlled  by the  parameter
  \begin{equation}D=2R_c-(\alpha_r-\alpha_p).\end{equation}
   Following to \cite{29},  it  has been shown that   the  parameter $R_c$ of shadows   takes   the following form
\begin{equation}
\label{d1}
R_c=\frac{(\alpha_t-\alpha_r)^2+\beta_t^2}{2|\alpha_t-\alpha_r|}.
\end{equation}
 The  distortion parameter  is characterized by  the ratio  of  $D_c$ and $R_c$
\begin{equation}
\label{delta}
\delta_c=\frac{D}{R_c}.
\end{equation}
These geometrical quantities will be discussed graphically. The associated calculations are
  illustrated in Fig.\ref{diso}.

One observes from the figure that, at fixed  brane number $N$, the size of the  shadow $R_c$ decreases by the increase of
 the rotation  parameter $a$.
 For the   distortion parameter, however,  $\delta_c$ increases by increasing  the rotation parameter  $a$.
  It is also noted  that  $\delta_c$ is almost null for $N\simeq 3$.
  When fixing the rotation $a$,  the quantity $R_c$ increases  with the D3-brane number to
   reach a maximum and then  it decreases.
   However,   the distortion parameter $\delta_c$ decreases to reach a vanishing value then it  increases gradually within increasing $N$. As it is shown in the previous section,  both parameters $R_c$ and $\delta_c$ are not always defined and are sensible to a selective value of the moduli space spanned by the triplet $(M,N,a)$.

\subsubsection*{Energy emission rates}

\begin{figure}[!htb]
\begin{center}
\centering
\begin{tabbing}
\centering
\hspace{7.9cm}\=\kill
\includegraphics[scale=.5]{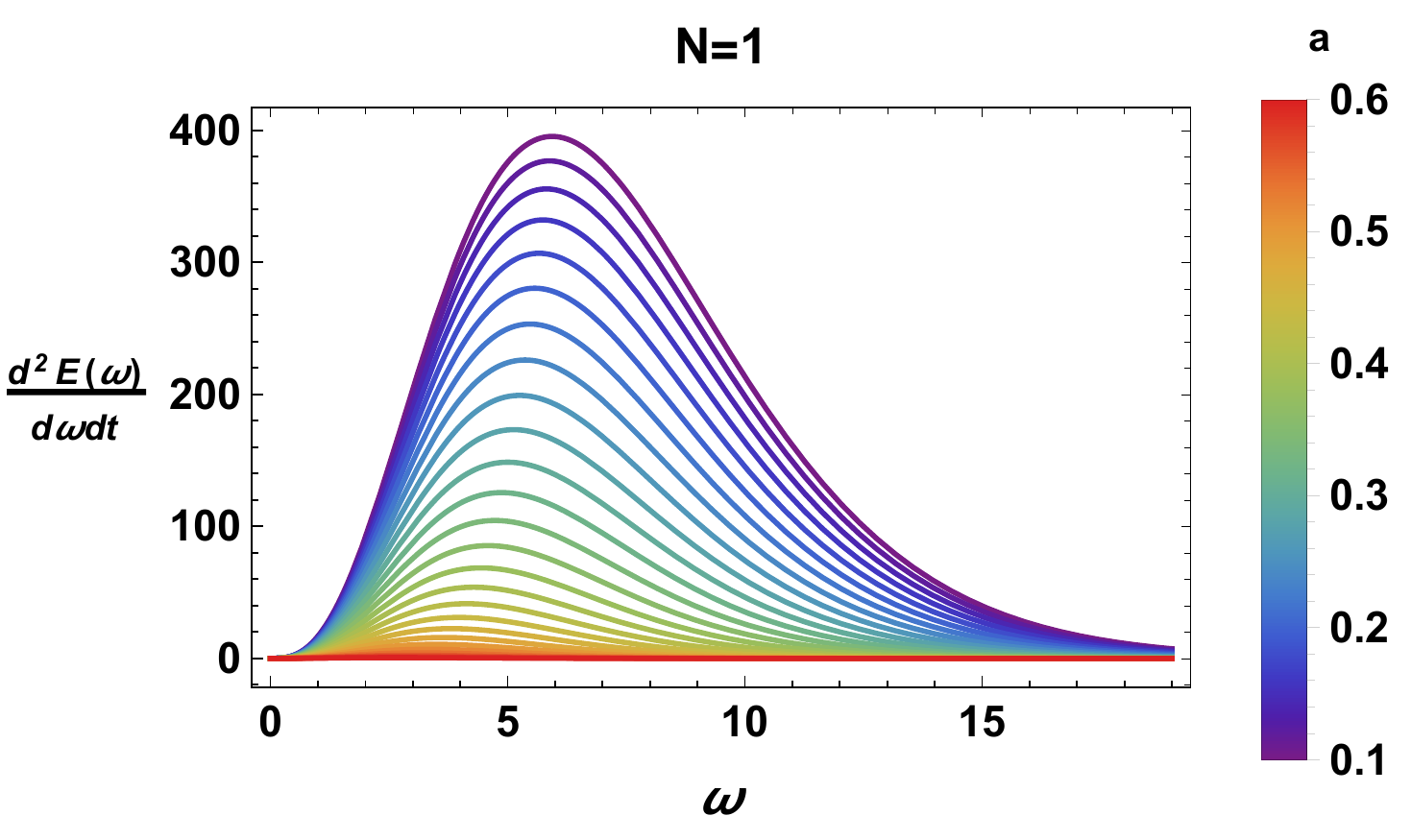} \>
\includegraphics[scale=0.5]{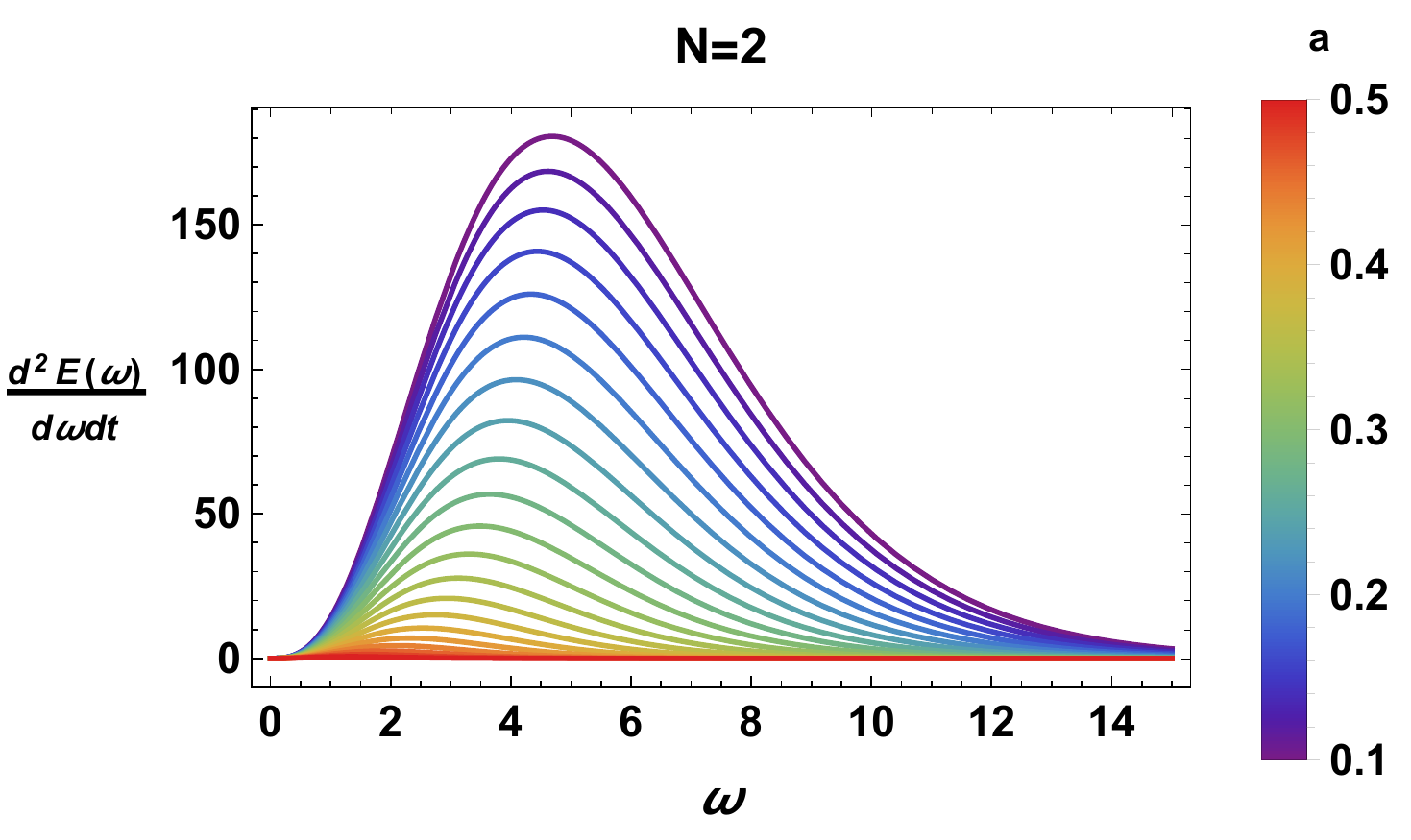} \\
\includegraphics[scale=0.5]{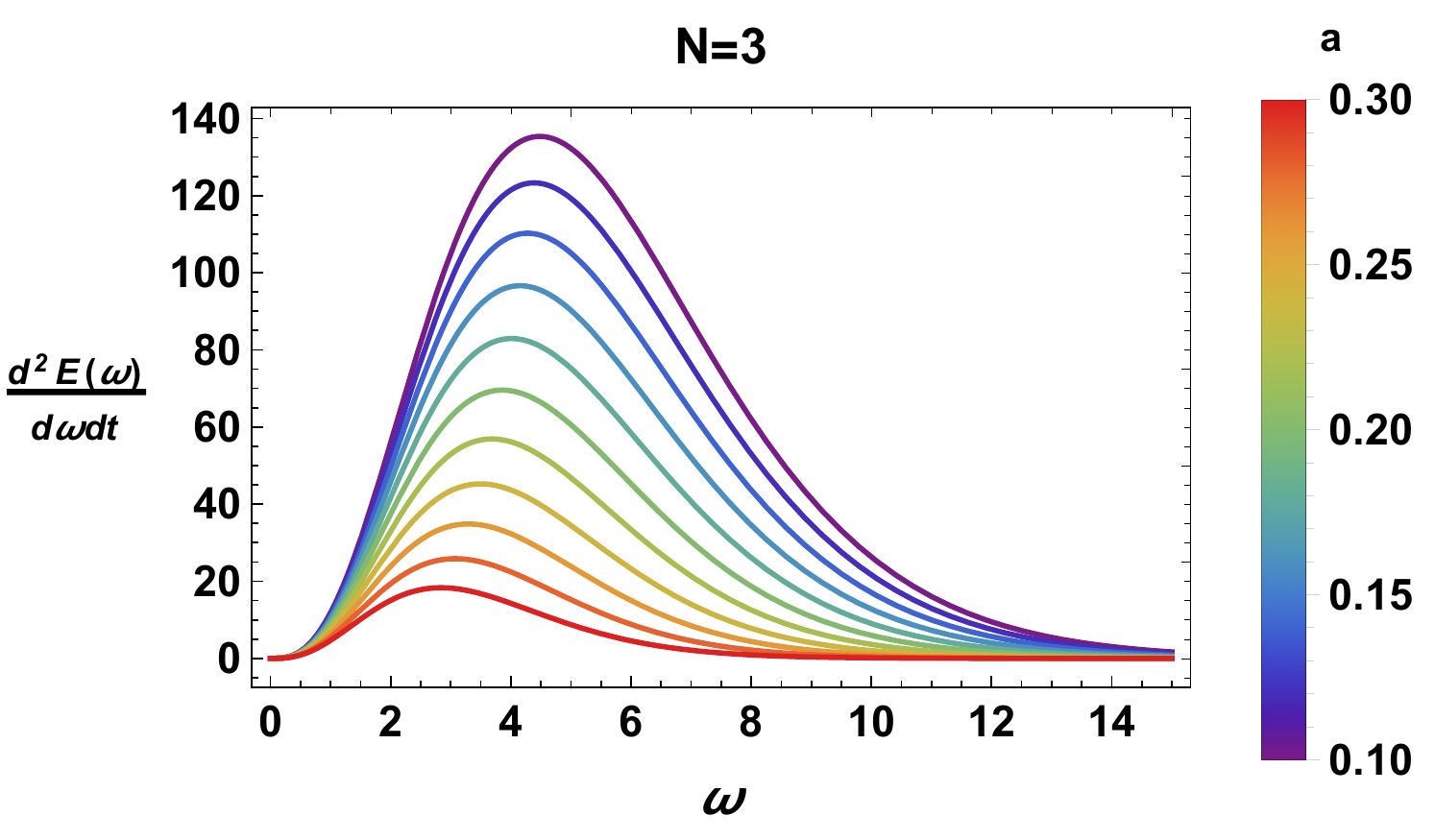} \>
\includegraphics[scale=.5]{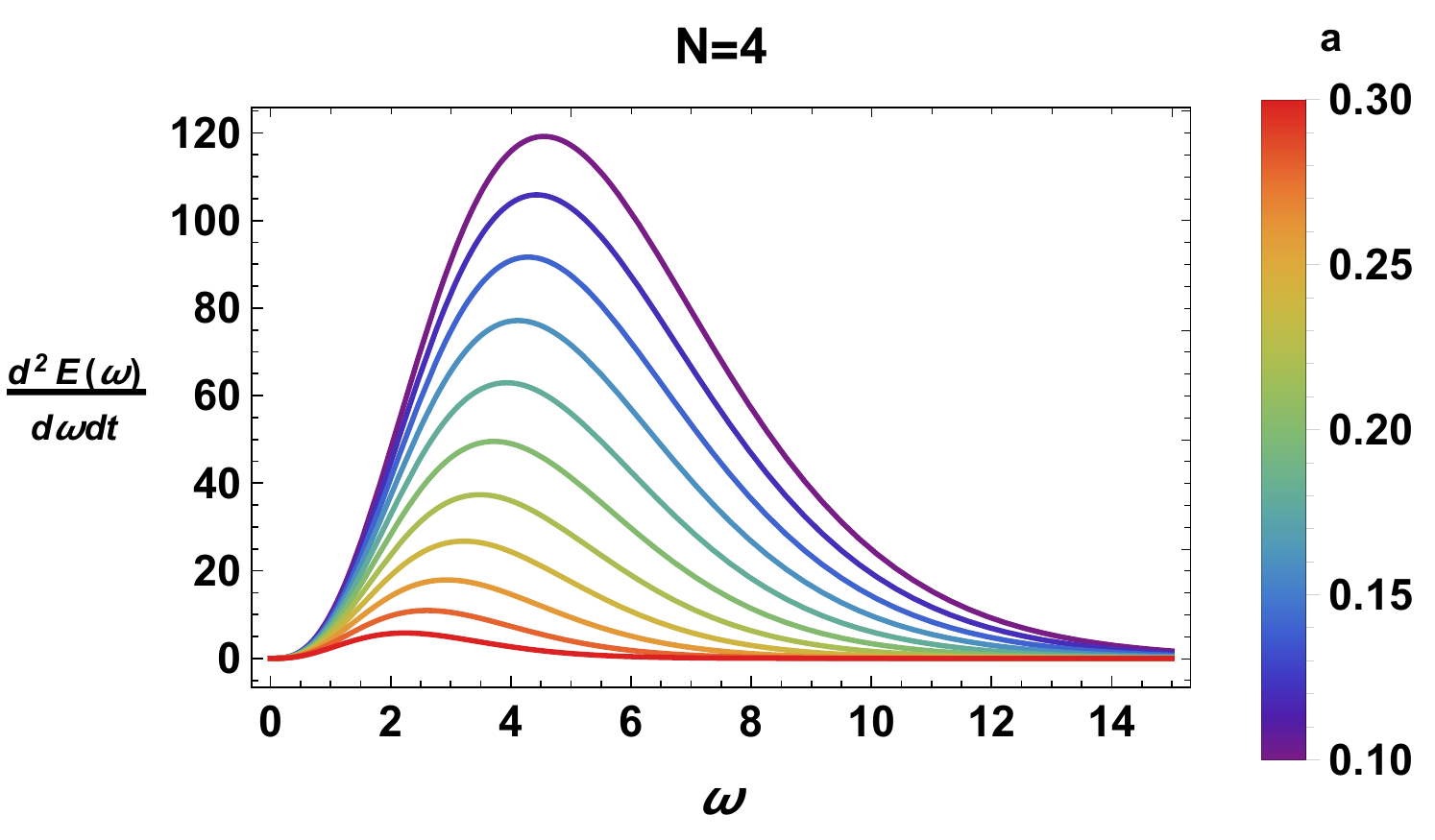} \\
\end{tabbing}
\caption{ Emission rate of  the  $  AdS_{5} \times \mathbb{S}^{5}$ black hole for different values of the rotation parameter $a$ and the brane numbers $N$, for $\ell_p=1$ and $M=1$.}
\label{shfa1}
\end{center}
\end{figure}

Here, we focus on   the  energy emission rate. Indeed,  for a far distant observer the high energy absorption cross section approaches to  the geometrical optical limit, the black hole shadow.  In intermediate
 regimes, the  absorption cross section of the black hole oscillates around limiting constant value $\sigma_{lim}$ at very high energy\cite{30,31}. This  limiting constant value,   being   approximately equal to the size of ``photon'' 3-sphere. In this approximation,  the energy emission rate can be written as
\begin{equation}
\frac{d^2 E(\omega)}{d\omega dt}=\frac{2\pi^2 \sigma_{lim}}{e^{\frac{\omega}{T_{{out}}}}-1}\omega^{4},
\end{equation}
where $\omega$ is the emission frequency, $R_{c}$ is the shadow radius and $T_{out}$ is the  Hawking temperature of the
  black hole.  In five dimensional space-time,    it is recalled that $\sigma_{lim}$ is approached as 
 \begin{equation}
 \sigma_{lim}\approx \frac{\pi^{3/2}R_c^{3}}{\Gamma(\frac{5}{2})}\end{equation}
   providing
\begin{equation}
\frac{d^{2} E(\omega)}{d \omega \, dt}= \frac{8 \pi^{3} \,  \, R^{3}_{c}}{3(e^{\frac{\omega}{T_{out}}}-1)}\omega^{4}.
\end{equation}
In this case,  the temperature $T_{out}$ can be expressed  as
\begin{equation}
T_{out}= \frac{r_h}{ \pi (r_{h}^{2}+a^{2})}  \left(1 + \frac{ (2r_{h}^{2}+a^{2})}{ 2^{1/4}\ell_{p}^{2} \, \sqrt{N}} \right),
\end{equation}
where $r_h$  is  the horizon radius.
We show the illustrate the behavior of the  energy emission rate versus the frequency for different values of the  rotation parameter and  the brane  number in Fig.\ref{shfa1}.

One observes from the formula, and is graphically clear in the figure, that the energy emission rate
varies only slowly.
The peak of  the energy emission rate  slowly shifts to lower frequencies and its height decreases  with the increase in the values of the  rotation parameter.  This behavior  is similar to what happens
to black holes embedded in  M-theory  compactifications\cite{27}. Finally we note that
increasing the  D3-brane number, the  energy  emission rate vanishes   by increasing the rotating parameter.

\section{ Further discussions and conclusions}

In this work,  we  have investigated  the shadow properties  of  5D  black holes embedded in type IIB superstring/supergravity inspired   spacetimes.
In particular,  we   have studied the effect of the presence of an arbitrary number of D3-branes  and of  the  rotation
on the shape  and    size of  the shadows.
For  non-rotating solutions, the circular size of the shadow decreases with the D3-brane number.
Larger values of  the rotation  parameter   decrease, however,   the size of the shadows.
This behavior  matches the  ``intuitive''  idea  indicating  that when  the black hole  rotates  rapidly,  null geodesics or
``photon'' orbits become  closer due to a   decreasing  in the effective gravitational effects.
After that, we have analyzed the geometrical observables associated with  the distortion and the size of the shadows, respectively.
Among others, we  have found  that the shapes   are  significantly  more and more distorted and the size is decreasing with the D3-brane number and the rotating parameter.  The energy emission rate aspects  of such 5D  black holes    have been  examined 
 in some details. It has been remarked that  the obtained  behavior is similar to the one observed in M-theory black holes
\cite{27}.
This  work opens up for further studies.
One interesting  question is to complete this analysis by considering  the general case where
the two rotating parameters supported by the $SO(5)$ Lie  symmetry are switched on.
In particular, it is interesting to write down  the associated shadow geometries. A natural question concerns the extension of this work to the case of  external sources  involving either  dark  energy  or  dark matter. We hope to report elsewhere on these open questions in  future \cite{newadil}.

\section*{Acknowledgments}
 AB would like to thank the Departamento de F\'isica, Universidad de Murcia for very kind hospitality.
The authors would like to thank N. Askor,   A. El Balali, S-E. Ennadifi,  P. Diaz,  J. J. Fern\'andez-Melgarejo,  M. P.  Garcia del Moral,  Y. Hassouni, K. Masmar,  M. B. Sedra and A. Segui for discussions on related topics.
The work of ET  has been  supported in part by the Spanish Ministerio de Universidades and Fundacion
Seneca (CARM Murcia) grants FIS2015-3454, PI2019-2356B and the Universidad de Murcia project E024-018.
This work is partially supported by the ICTP through AF-13.

\end{document}